\documentclass[preprint,12pt]{elsarticle}
\makeatletter
\def\ps@pprintTitle{%
	\let\@oddhead\@empty
	\let\@evenhead\@empty
	\def\@oddfoot{\hfill}
	\let\@evenfoot\@oddfoot}
\makeatother

\usepackage{amssymb}
\usepackage{graphicx}
\usepackage{float}
\usepackage{siunitx}
\usepackage{amsmath}\usepackage{graphicx}%
\usepackage{multirow}%
\usepackage{amsmath,amssymb,amsfonts}%
\usepackage{amsthm}%
\usepackage{mathrsfs}%
\usepackage[title]{appendix}%
\usepackage{xcolor}%
\usepackage{textcomp}%
\usepackage{manyfoot}%
\usepackage{booktabs}%
\usepackage{algorithm}%
\usepackage{algorithmicx}%
\usepackage{algpseudocode}%
\usepackage{listings}%
\usepackage{float}
\usepackage{siunitx}
\usepackage{soul}
\usepackage[margin=1in]{geometry}

\sethlcolor{white}
\graphicspath{{./figs/}}
\begin{document}
\begin{frontmatter}
\title{Plastic Work Partitioning During Slip- and Twinning-Dominated Deformation in AZ31B Magnesium Alloy}
%
\author[a]{Micha{\l} Maj\corref{cor1}}
\author[a]{Sandra Musia{\l}}
\author[a]{Marcin Nowak}
%
\address[a]{Institute of Fundamental Technological Research, Polish Academy of Sciences, Pawińskiego 5B, 02-106 Warsaw, Poland}

\cortext[cor1]{Corresponding author: Micha{\l} Maj, mimaj@ippt.pan.pl}

\begin{abstract}
	
Plastic work partitioning was investigated in extruded AZ31B magnesium alloy under loading orientations promoting slip- or twinning-dominated deformation. Slip-dominated deformation exhibits stable plastic flow with approximately 50$\%$ of plastic work dissipated as heat. \textcolor{black}{But} twinning-dominated deformation initially stores most plastic work, delaying dissipation and promoting rapid strain hardening and early strain localization. \textcolor{black}{These} distinct mechanical responses correlate with microstructural evolution, \textcolor{black}{showing that, in HCP magnesium alloys, the energy storage depends on the active deformation mechanism}.
\end{abstract}


\end{frontmatter}

Due to its high specific strength, AZ31B magnesium alloy is widely used in the automotive and aviation industries. Its hexagonal close-packed (HCP) crystal structure results in strong elastic-plastic anisotropy when the material is deformed along different directions. The macroscopic mechanical response differs significantly when loading is applied parallel or transverse to the unit cell c-axis, which is associated with the activation of different \textcolor{black}{mechanisms of plastic deformation}: mechanical twinning dominates when \textcolor{black}{the} loading is parallel to the c-axis, whereas dislocation slip \textcolor{black}{dominates when the loading is transverse} \cite{wang2014,knezevic2010}. Among the various twinning modes, the most prevalent in magnesium alloys are $\{10\bar{1}2\}\langle10\bar{1}1\rangle$ extension twinning and $\{10\bar{1}1\}\langle10\bar{1}2\rangle$ contraction twinning, activated by tensile and compressive loading along the c-axis, respectively \cite{barnett2007,sun2014}. AZ31B alloy, while polycrystalline in industrial applications, typically develops a strong fiber texture during processing. After extrusion, for instance, the c-axes of most grains are oriented nearly \textcolor{black}{perpendicularly} to the extrusion direction (ED). Consequently, the mechanical response becomes highly anisotropic, \textcolor{black}{depending on whether it is tested along the ED} or transverse to the ED. Dislocation slip is readily activated during loading along the ED, \textcolor{black}{but} mechanical twinning is more likely to occur under transverse loading \cite{yi2006,jiang2007,frydrych2020}. The competition between dislocation slip and deformation twinning not only governs \textcolor{black}{the} anisotropy but also \textcolor{black}{greatly influences the} strain hardening, strain localization, and fracture behavior. A quantitative understanding of how these mechanisms accommodate plastic strain is therefore essential for predicting the mechanical response of textured magnesium alloys.

Although the mechanical behavior of AZ31B alloy has been extensively characterized, the thermodynamic aspects of \textcolor{black}{its} plastic deformation\textemdash specifically the partitioning of plastic work into stored energy and heat\textemdash remain insufficiently \textcolor{black}{understood}. This partitioning is directly linked to strain hardening, defect accumulation, and the onset of strain localization. A major challenge lies in the high thermal diffusivity of the alloy ($\alpha\approx$ \,\qty{55}{mm^2/s}, nearly 16 times higher than that of austenitic steel, $\alpha\approx$ \,\qty{3.33}{mm^2/s}), which substantially complicates \textcolor{black}{a} thermomechanical analysis.

Few studies have investigated the energy dissipation and storage during plastic deformation of AZ31B alloy, and most \textcolor{black}{of those that have, have focused on the} dynamic loading, where heat-transfer effects are often neglected \cite{ghosh2017, kingstedt2019}. Under such conditions, accurate temperature measurements via infrared thermography (IRT) are challenging, which may \textcolor{black}{affect the accuracy} of the recorded data.
In References \cite{ghosh2017} and \cite{kingstedt2019}, a split-Hopkinson pressure bar was used for the mechanical testing, and a liquid-nitrogen cooled HgCdTe detector was used to record infrared radiation from a fixed \qty{100}{\micro\metre}×\qty{100}{\micro\metre} region. During deformation, the specimen moves relative to this window, so the measured temperature \textcolor{black}{is for} different material points over time. Comparing the local dissipated energy with global mechanical work assumes uniform deformation across the gauge section\textemdash \textcolor{black}{a condition that may become difficult to satisfy under dynamic loading and therefore requires careful validation.}
Moreover, \cite{rittel2007}, the primary reference for this method, \textcolor{black}{notes that a rigorous quantification of measurement accuracy remains challenging.} In addition, straightforward estimates based on the data reported in Reference \cite{ghosh2017} indicate that \textcolor{black}{the estimated temperature increase may differ substantially from the theoretical upper bound, in some cases by up to approximately 50\%}.
In contrast, \cite{ozdur2025} studied AZ31B alloy under quasi-static, non-monotonic loading ($\dot{\varepsilon} = 2 \cdot 10^{-3}~\si{s^{-1}}$). The reported maximum temperature changes ($\approx$ \qty{100}{mK}) were 3.5 times smaller than the thermal sensitivity of the bolometric system used ($\approx$ \qty{350}{mK}), \textcolor{black}{indicating that the measured temperature variations are close to the sensitivity limit of the system, such that the quantitative results may be significantly affected by signal-processing procedures and measurement uncertainty.} 
Thus, \textcolor{black}{previous} conclusions\textemdash e.g., that energy dissipated by slip exceeds that for twinning\textemdash \textcolor{black}{still require further experimental validation}. The present study quantitatively evaluates \textcolor{black}{the partitioning of the plastic work} in extruded AZ31B under loading orientations promoting either slip- or twinning-dominated deformation. By combining full-field strain measurements with spatially resolved temperature analysis and accounting for anisotropic thermal conductivity, \textcolor{black}{the present study provides a direct and experimentally robust assessment of the coupling between deformation mechanisms, mechanical response, and energy storage in an HCP magnesium alloy. This approach delivers quantitative experimental evidence supporting trends previously predicted by theoretical studies.}
The fraction of plastic work converted to heat by slip- and twinning-mediated deformation was quantified using the transient heat-conduction-based approach proposed in Reference \cite{musial2022}, adapted for the anisotropic thermal properties of the extruded Mg alloy. In the first step, the components of \textcolor{black}{the} thermal conductivity tensor were determined using the solution of the heat conduction equation for a semi-infinite body with Neumann boundary conditions (constant heat flux $q_0$ on the surface) (Eq. \ref{eq:cond_comp}).

\begin{equation}
	\label{eq:cond_comp}
	T(x,t) - T_0 =
	\frac{2 q_0}{k}
	\sqrt{\frac{\alpha t}{\pi}}
	\exp\left(-\frac{x^2}{4 \alpha t}\right)
	-
	\frac{q_0 x}{k}
	\operatorname{erfc}\!\left(\frac{x}{2\sqrt{\alpha t}}\right),
\end{equation}
where $T(x,t)$ is the temperature at depth $x$ and time $t$, $T_0$ the initial temperature, $k$ the thermal conductivity, $\alpha = {k}/{\rho c}$ the thermal diffusivity, $\rho$ the density, $c$ the specific heat capacity, and $\operatorname{erfc}(\cdot)$ the complementary error function.

A \qty{16}{mm} cubic specimen cut from a \qty{25}{mm} diameter extruded rod was placed on a massive plate, with a constant-power Peltier module on its top surface. A thin layer of eutectic GaLiSn ensured good thermal contact at both interfaces. A Dirichlet boundary condition ($T(t)$ = $T_0$) at the bottom face extended the validity of the semi-infinite approximation, even for \textcolor{black}{material with a high thermal diffusivity}.
The evolution of \textcolor{black}{the} temperature distribution during heating with \textcolor{black}{a} constant heat flux was measured using \textcolor{black}{a} Phoenix IR camera with taking frequency \qty{346}{Hz} and thermal sensitivity \qty{20}{mK}. The \textcolor{black}{evolutions of the temperatures at six points,} placed at: \qty{0.77}{mm}, \qty{1.33}{mm}, \qty{1.88}{mm}, \qty{2.44}{mm}, \qty{2.99}{mm}, \qty{3.54}{mm} from the heated surface, were simultaneously fitted to \textcolor{black}{the} analytical solution (see Eq.\ref{eq:cond_comp}). The procedure was repeated for all three directions of the cube and $k$ values were determined: \( k_{\parallel} = 95.35 \pm 0.60\ \mathrm{W\,m^{-1}K^{-1}} \), \( k_{\perp1} = 87.02 \pm 0.52\ \mathrm{W\,m^{-1}K^{-1}} \) and \( k_{\perp2} = 87.12 \pm 0.48\ \mathrm{W\,m^{-1}K^{-1}} \). Thus, the values \(k_{\parallel} = 95\ \mathrm{W\,m^{-1}K^{-1}} \) and the \(k_{\perp} = 87\ \mathrm{W\,m^{-1}K^{-1}} \) were used in \textcolor{black}{the subsequent} calculations.

\begin{figure}[H]
	\centering
	\includegraphics[width=13.6cm]{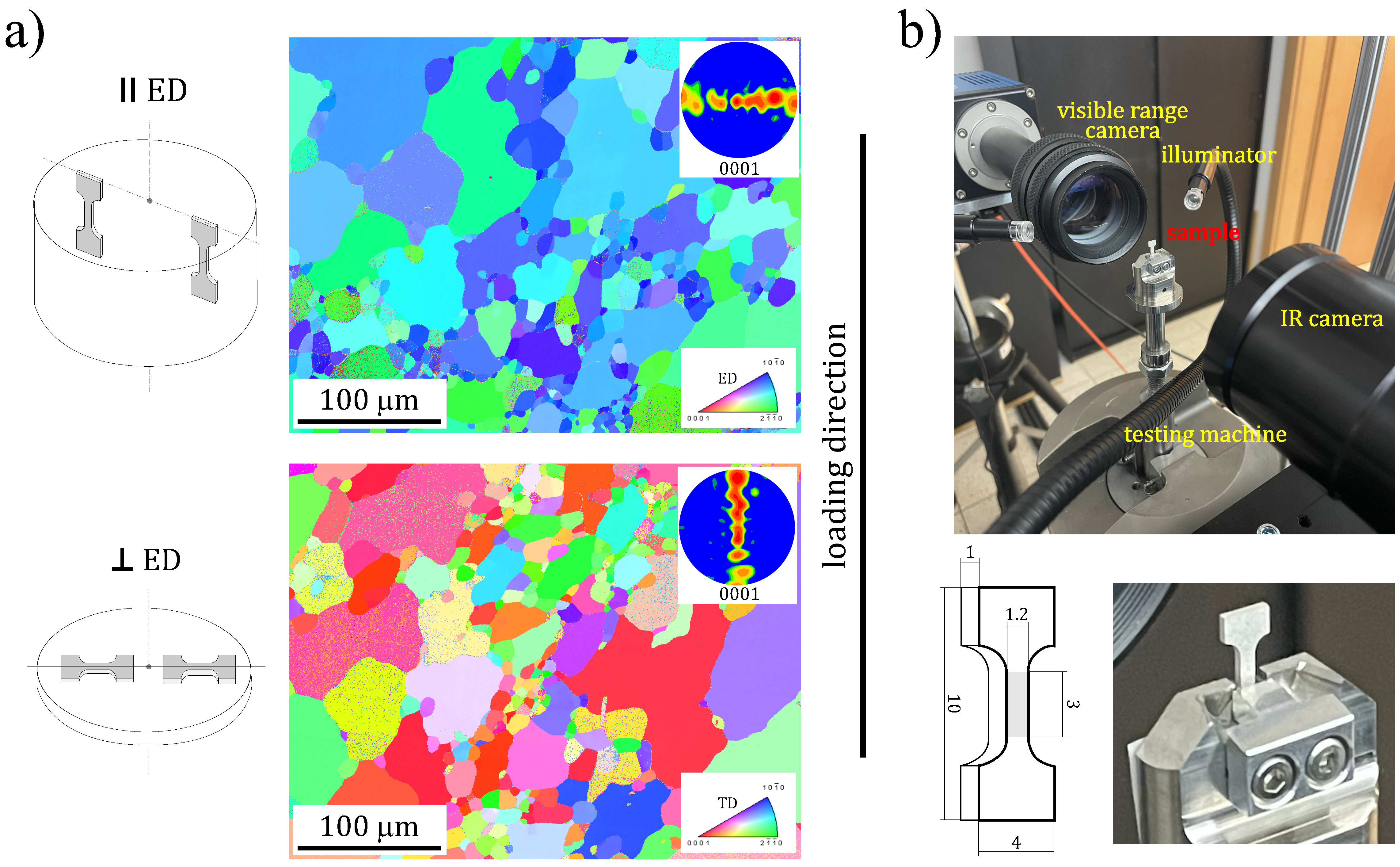}
	\caption{Specimen orientation and initial microstructure (a); experimental setup for plastic work partitioning measurements and specimen geometry (b).}
	\label{fig1}
\end{figure}

Two types of specimens were prepared to promote distinct dominant deformation mechanisms: dislocation slip (specimen $\parallel$\,ED) and deformation twinning (specimen $\perp$\,ED), as illustrated in Fig.~\ref{fig1}(a). The initial microstructures of both \textcolor{black}{types of specimen} are also shown as inverse pole figure (IPF) maps projected onto the loading direction for the subsequent tensile tests.\\
The experimental setup, enabling spatially resolved correlation of local plastic strain and internal heat sources and allowing \textcolor{black}{an analysis of the energy} partitioning under heterogeneous deformation, is shown in Fig.~\ref{fig1}(b). The main components include a ThermaCam Phoenix infrared camera, a PCO 5.5 visible-range camera, and an MTS 858 hydraulic testing machine. The main settings of the optical devices used in the experiments are listed in Table \ref{tab:tab1}. \textcolor{black}{The noise-equivalent temperature difference (NETD) and systematic errors of the thermographic system were determined using a TCB 4D blackbody with microscopic optics and an integration time of \qty{3}{ms}, consistent with the experimental conditions. The obtained NETD values were \qty{39}{mK} at \qty{25}{\celsius} and \qty{40}{mK} at \qty{30}{\celsius}, while the corresponding mean/maximum systematic errors were \qty{0.039}{\celsius}/\qty{0.054}{\celsius} and \qty{0.041}{\celsius}/\qty{0.055}{\celsius}, respectively.}
The specimens with surfaces prepared for DIC and IRT analyses, were subjected to displacement-controlled tension at a displacement rate of \qty{2}{mm/s}, which corresponds to a strain rate $\dot{\varepsilon} = 6.66 \cdot 10^{-1} \text{ s}^{-1}$, taking into account the \textcolor{black}{geometry of the specimen}. \textcolor{black}{This strain rate was selected on the basis of preliminary experiments as a compromise between mechanical and thermal requirements. The stress–strain response was found to be essentially rate-independent relative to quasi-static conditions $(\dot{\varepsilon} = 1 \cdot 10^{-3} \text{ s}^{-1})$, while still producing temperature changes large enough to be resolved by the infrared measurement system. Three independent experiments were performed for each loading orientation, for which stress–strain curves and the corresponding evolutions of the mean temperature were determined, demonstrating very high repeatability. The mean and maximum standard deviations of the stress–strain curves were \qty{1.21}{MPa} and \qty{8.66}{MPa} for the $\perp$\,ED specimens, and \qty{1.06}{MPa} and \qty{3.36}{MPa} for the $\parallel$\,ED specimens, respectively. The corresponding standard deviations of the mean temperature were \qty{0.016}{K} and \qty{0.073}{K} for the $\perp$\,ED specimens, and \qty{0.023}{K} and \qty{0.069}{K} for the $\parallel$\,ED specimens, respectively.}

\begin{table}[h]
	\caption{Visible range and infrared camera parameters used in experiments}
	\label{tab:tab1}
	\centering
	\begin{tabular}{@{}llll@{}}
		\toprule
		\multicolumn{2}{c}{PCO Edge 5.5} & \multicolumn{2}{c}{ThermaCam Phoenix} \\
		\cmidrule(lr){1-2} \cmidrule(lr){3-4}
		Parameter & Value & Parameter & Value \\
		\midrule
		Resolution        & 1280$\times$470 pixels  & Resolution           & 320$\times$256 pixels \\
		Binning           & 2$\times$2              & Thermal sensitivity  & \qty{20}{mK} \\
		Pixel size        & \qty{8.5}{\micro\metre} & Pixel size           & \qty{13.9}{\micro\metre} \\
		Exposure time     & \qty{3}{ms}             & Integration time     & \qty{3}{ms} \\
		Taking frequency  & \qty{100}{Hz}           & Taking frequency     & \qty{100}{Hz} \\
		\bottomrule
	\end{tabular}
\end{table}

\begin{figure}[H]
	\centering
	\includegraphics[width=12.0 cm]{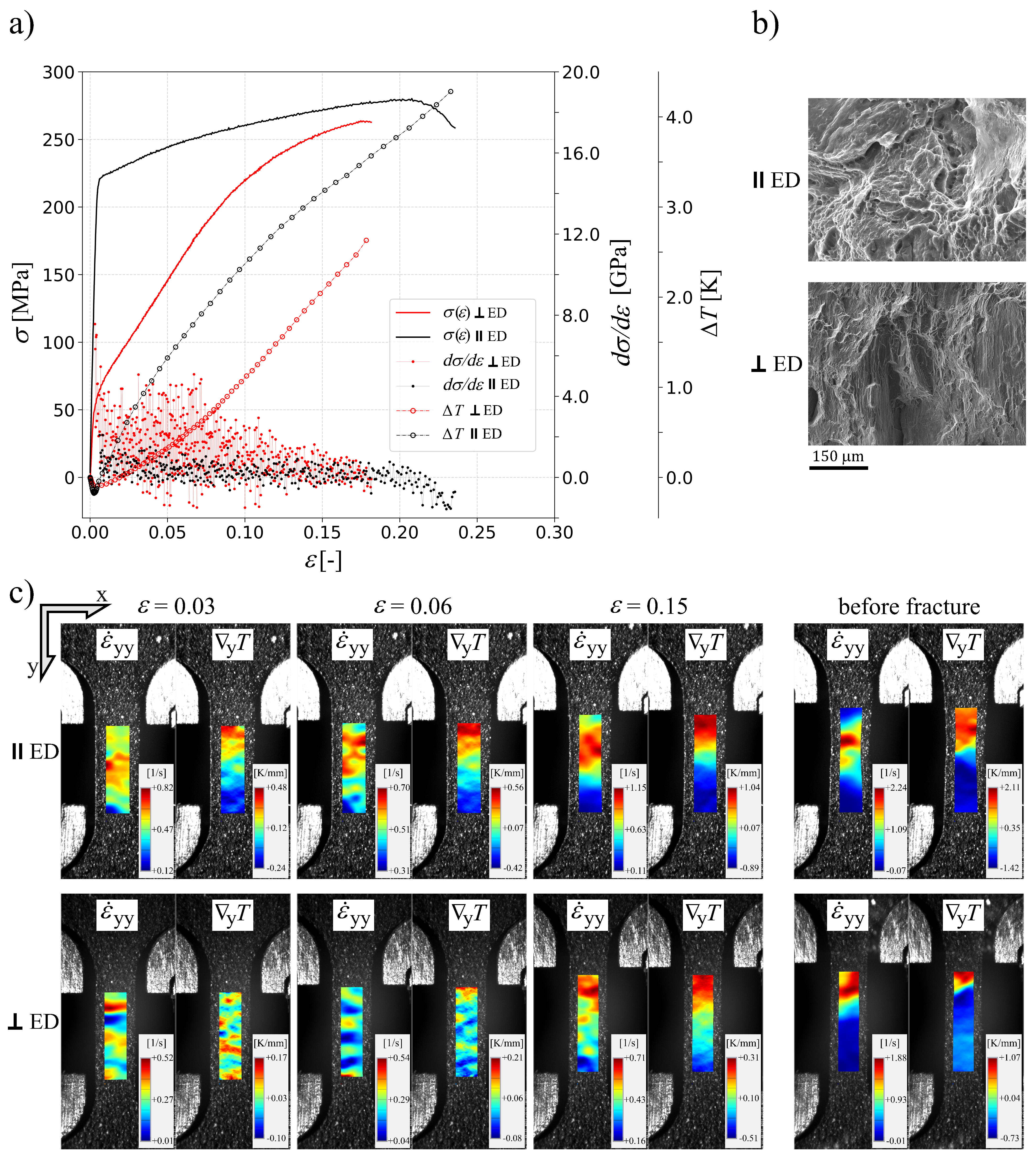}
	\caption{Stress-strain curves for slip-dominated ($\parallel$\,ED) and twinning-dominated ($\perp$\,ED) loading directions, with the corresponding strain hardening coefficients ${d\sigma}$/${d\varepsilon}$ and mean temperature changes $\Delta T_{mean}$ (a). Corresponding fracture surfaces (b) and the distributions of the axial component of Hencky strain rate $\dot{\varepsilon}_{yy}$, and temperature gradient $\nabla_{y} T$ (c).}
	\label{fig2}
\end{figure}

Based on the image sequences recorded simultaneously during \textcolor{black}{the} straining, the evolutions of the temperature field (IR camera) and the displacement field (visible-range camera and Digital Image Correlation (DIC) algorithm) were determined and coupled, as described in Reference \cite{nowak2018}. \textcolor{black}{A} DIC analysis was performed using circular subsets (25 px), Zero-Normalized Cross-Correlation with bicubic interpolation, yielding approximately 32000 measurement points per frame. \textcolor{black}{The full-field analysis was performed on one specimen of each type.} \textcolor{black}{The uncertainty of temperature and displacement measurements was evaluated from sequences of 100 images acquired under testing conditions prior to mechanical loading. The mean temperature measurement uncertainties were \qty{0.0029}{\celsius} and \qty{0.0027}{\celsius}, for the $\parallel$\,ED and $\perp$\,ED specimens, respectively. Moreover, the uncertainties associated with numerical differentiation of the temperature field were determined and were equal to \qty{0.0218}{K/s} for the time derivative $\dot{T}$ and \qty{0.0015}{K/mm} for the temperature gradient $\frac{\partial T}{\partial y}$. The corresponding uncertainties of the measurement of displacement components were as follows: for the $\parallel$\,ED: $u_x$=\qty{0.063}{\micro\metre}, $u_y$=\qty{0.054}{\micro\metre}; for the $\perp$\,ED: $u_x$=\qty{0.042}{\micro\metre}, $u_y$=\qty{0.087}{\micro\metre}.}
\textcolor{black}{The accuracy of the displacement gradients determination was evaluated using a hybrid experimental–synthetic validation procedure, in which experimentally acquired images were numerically deformed using prescribed analytical shape functions, as described in detail in the Appendix of Reference \cite{musial2022}. The maximum and mean relative errors were 0.032 and 0.015 for the $\parallel$\,ED specimen and 0.033 and 0.017 for the $\perp$\,ED specimen, respectively.}\\
The surface distribution of the power density of \textcolor{black}{the} heat sources $\dot{q}_v$ was obtained from the \textcolor{black}{two-dimensional} transient heat conduction equation \textcolor{black}{(justified by the specimen dimensions and the high thermal diffusivity of the alloy)} formulated to account for the anisotropic thermal properties of the \textcolor{black}{investigated} material:
\begin{equation}
	\rho c\,\dot{T}
	=
	k_x \frac{\partial^2 T}{\partial x^2}
	+
	k_y \frac{\partial^2 T}{\partial y^2}
	+
	\dot{q}_v\;\left[\, \si{W/mm^3} \right],
	\label{eq:heat_cond_eq}
\end{equation}
where $\dot{T}$=${\partial T}/{\partial t}$+$\nabla T \cdot \mathbf{v}$ is the temperature material time derivative; $k_x$ and $k_y$ the thermal conductivity in $x$ and $y$ directions, respectively; and $\mathbf{v}$ the material velocity field. Based on the thermal conductivity components $k_{\perp}$ and $k_{\parallel}$  derived earlier for the specimen strained $\parallel$ ED the values of $k_x$ and $k_y$
were taken as \qty{87}{W\,m^{-1}K^{-1}} and \qty {95}{W\,m^{-1}K^{-1}}, respectively, whereas for the specimen strained $\perp$ ED both $k_x$ and $k_y$
were taken as \qty{87}{W\,m^{-1}K^{-1}}.\\
In Figure \ref{fig2}, the stress-strain curves for specimens strained parallel and perpendicular to the ED are presented. The mechanical response differs markedly between the two orientations, reflecting the dominant deformation mechanism activated under each loading condition. The total deformation energy, represented by the area under the curve, is significantly higher for the slip-dominated specimen (\(\parallel\)~ED) than for the twinning-dominated one (\(\perp\)~ED). The slip-dominated specimen also exhibits a substantially higher yield stress, indicating that a larger applied load is required to initiate plastic deformation when dislocation glide is the primary deformation mechanism. The elevated strain hardening observed in the twinning-dominated specimen is consistent with progressive twin formation and associated barriers to dislocation motion. This leads to a reduced dislocation mean free path and enhanced dislocation accumulation, commonly described as a \emph{dynamic Hall-Petch like effect} \cite{allain2004,gutierrez2011}. Continued twin nucleation and growth sustain a high strain-hardening rate over a wide \textcolor{black}{range of strain}, whereas in the slip-dominated case the evolution toward quasi-stable dislocation structures results in a gradual \textcolor{black}{reduction of hardening}. At large strains, the $\parallel$\,ED specimen exhibits a gradual load decrease associated with diffuse necking prior to failure, while the $\perp$\,ED specimen fails abruptly without \textcolor{black}{a} pronounced load drop, consistent with a more brittle fracture mode, as confirmed by the fracture surfaces in Fig. \ref{fig2}(b).\\
The temporal evolution of the ${d\sigma}$/${d\varepsilon}$ coefficient for the $\perp$\,ED specimen exhibits a larger scatter in the strain range dominated by twinning (up to a strain of approximately 0.07) compared with the predominantly slip-controlled regime observed at higher strains for $\perp$\,ED and throughout most of the deformation process of the $\parallel$\,ED specimen. Such \textcolor{black}{a} scatter, an inherent feature of twinning, is particularly pronounced in single crystals; although attenuated in polycrystals, it remains detectable \cite{salem2003}. With increasing strain, the scatter progressively decreases, and at the final stages of deformation of \textcolor{black}{the} $\perp$\,ED specimen it becomes comparable to that of the slip-dominated regime. 

The same plot also shows the evolution of the mean temperature change \(\Delta T_{\mathrm{mean}}\). The $\parallel$ ED orientation exhibits markedly higher \textcolor{black}{values of} \(\Delta T_{\mathrm{mean}}\), exceeding \qty{4}{K}, whereas the $\perp$ ED orientation reaches only slightly above \qty{2.5}{K}. The corresponding maximum temperature \textcolor{black}{increases} were \qty{5.38}{K} and \qty{3.62}{K}, respectively, over a total process duration of approximately \qty{0.5}{s}.\\
Figure \ref{fig2}(c) presents the distributions of the Hencky strain rate component $\dot{\varepsilon}_{yy}$ and the corresponding temperature gradients $\nabla_{y} T$  along the loading direction.
In the twinning-dominated specimen, deformation localizes into narrow macroscopic bands, more clearly revealed in the $\nabla_{y} T$ maps. \textcolor{black}{Nevertheless, owing to the rapid thermal diffusion in the material, the temperature field remains spatially continuous, enabling reliable determination of the corresponding heat sources using the adopted methodology. In contrast,} the slip-dominated specimen exhibits macroscopically more homogeneous deformation, consistent with the \textcolor{black}{evolution of}  ${d\sigma}$/${d\varepsilon}$ observed previously.

To interpret the observed differences in strain hardening and localization, the partitioning of plastic work into stored energy and heat was evaluated. To quantify the energy dissipated as heat, the power density of the heat sources $\dot{q}_v$ at each point on the surface was integrated over time, and the resulting mean heat dissipation in the gauge section was evaluated at each instant of straining and compared with the plastic work derived from the stress-strain curves. \textcolor{black}{The systematic error of the energy dissipated as heat was evaluated on the basis of the experimental data recorded prior to loading, for which a zero heat-generation value is expected. The resulting non-zero values were used to quantify the systematic error of the method. The mean and maximum systematic errors of $\Delta q_d$ were found to be \qty{0.026}{J/g} and \qty{0.058}{J/g}, respectively.} 
Figure \ref{fig3} shows the evolution of plastic work $w_p$, energy dissipated as heat $q_d$ and stored energy $e_s$ for specimens strained $\parallel$\,ED (a) and $\perp$\,ED (b) together with the corresponding integral $\beta_{int}$ and differential $\beta_{diff}$ Taylor\textendash Quinney coefficients (c). For slip-dominated deformation ($\parallel$\,ED), $w_p$ and $q_d$ increase nearly linearly, reaching approximately \qty{33}{J/g} and \qty{17.5}{J/g} at $\varepsilon^p\approx$ 0.23, indicating a stable conversion of plastic work into heat. The initial rapid \textcolor{black}{increase} of $\beta_{int}$, followed by a slower increase, reflects the progressive formation of \emph{low-energy dislocation structures (LEDS)} \cite{hansen1986,ma2021,jiang2014,oliferuk2004}.
In twinning-dominated deformation $\perp$\,ED $w_p$ and $q_d$ increase nonlinearly, reaching lower values ($\approx$ \qty{19}{J/g} and $\approx$ \qty{8}{J/g} at $\varepsilon^p\approx$ 0.18). Heat dissipation remains minimal up to $\varepsilon^p\approx$ 0.07, implying that most plastic work is stored, likely as interfacial energy of twin boundaries. The delayed onset of dissipation is consistent with progressive energy storage in twin interfaces and evolving defect structures, which contributes to increased strain hardening but promotes mechanical instability at lower strains. The small initial dissipation likely reflects the dislocation-mediated character of twinning in \textcolor{black}{HCP} metals \cite{he2020}, whereas its lower magnitude compared to slip-dominated deformation can be attributed to the \textcolor{black}{relatively low frictional resistance of coherent twin boundary motion. The pronounced fluctuations in $\beta_{diff}$ closely correlate with the scatter in the strain hardening coefficient $d\sigma/d\varepsilon$ (Fig. \ref{fig2}a), indicating that they are predominantly of physical rather than experimental origin. Such intermittent behavior is characteristic of twinning-mediated deformation and is consistent with the classical concept of the “twinning cry,” associated with rapid propagation of partial dislocations during twin growth \cite{cottrell1953,mahajan1973,christian1985}. With increasing strain, the fluctuation amplitude progressively decreases as deformation transitions from twinning- to slip-dominated plasticity, accompanied by a steady increase in energy dissipation.}
Overall, slip-dominated deformation directs a larger fraction of the plastic work into heat, whereas twinning-dominated deformation stores more energy in the microstructure, particularly at early stages of loading. \textcolor{black}{The cumulative} $\beta_{int}$ \textcolor{black}{increases} from $\approx$ 0.33 to $\approx$ 0.52 for $\parallel$\,ED but from near zero to $\approx$ 0.4 for $\perp$\,ED. The instantaneous coefficient $\beta_{diff}$ is 0.5\textendash 0.7 for slip-dominated deformation but varies widely\textemdash from near zero to $\approx$ 0.7\textemdash for the $\perp$\,ED, reflecting the evolving interplay of twinning and slip in partitioning plastic work between storage and dissipation.

\begin{figure}[H]
	\centering
	\includegraphics[width=11 cm]{Fig_3}
	\caption{Plastic work $w_p$, energy dissipated as heat $q_d$ and stored energy $e_s$ as functions of plastic strain for slip-dominated ($\parallel$\,ED) (a) and twinning-dominated ($\perp$\,ED) specimens (b), along with the corresponding evolution of the integral $\beta_{int}$ and differential $\beta_{diff}$ Taylor\textendash Quinney coefficients (c).}
	\label{fig3}
\end{figure}  

\begin{figure}[H]
	\centering
	\includegraphics[width=8.7cm]{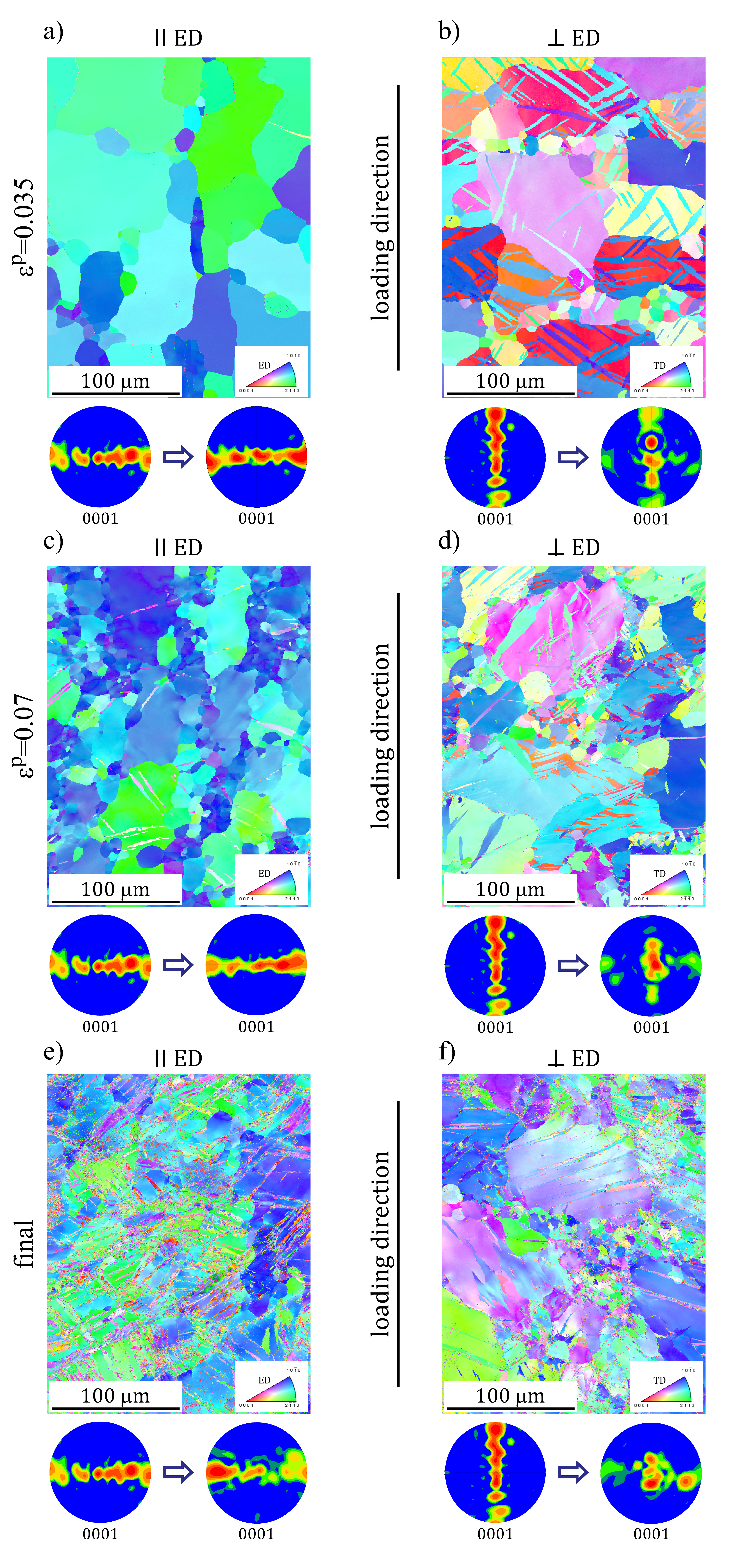}
	\caption{\textcolor{black}{EBSD maps and corresponding pole figures at different deformation stages ($\varepsilon^{p}$ = 0.035 (a-b), $\varepsilon^{p}$ = 0.07 (c-d), and final state (e-f)) for $\parallel$\,ED and $\perp$\,ED specimens.}}
	\label{fig4}
\end{figure}

To verify whether the inferred differences in energy storage and \textcolor{black}{dissipation} are \textcolor{black}{reflected in the evolution of the microstructure, the EBSD analyses were performed using Zeiss CrossBeam 350 scanning electron microscope equipped with Hikari Super camera, at early deformation stages ($\varepsilon^{p}$ = 0.035 and 0.07) and at the final stage (adjacent to the fracture) for both loading orientations. The strain levels $\varepsilon^{p}$ = 0.035 and 0.07 correspond to the regime in which twinning is expected to dominate in the $\perp$\,ED specimen. As shown in (Fig. \ref{fig4}(a-d)), a pronounced difference in twinning activity is observed between $\perp$\,ED and $\parallel$\,ED specimens. The twin boundary density, $\rho_{\mathrm{twin}}$, increases from $0.19$ to $0.31~\si{\micro\metre^{-1}}$ for the $\perp$\,ED specimen, while remaining negligible for the $\parallel$\,ED specimen ($0.002$ to $0.021~\si{\micro\metre^{-1}}$). This confirms that plastic deformation in the $\perp$\,ED orientation is strongly governed by twinning, whereas slip dominates in the $\parallel$\,ED case. The significantly higher twin boundary density in the $\perp$\,ED specimen provides direct evidence that a substantial portion of plastic work is stored as interfacial energy of twin boundaries. This observation is consistent with the low initial Taylor\textendash Quinney coefficient and delayed heat dissipation observed for this orientation. In contrast, the negligible twinning activity in the $\parallel$\,ED specimen supports the interpretation that energy storage is primarily associated with dislocation accumulation, resulting in a more continuous conversion of plastic work into heat.}
The orientation maps adjacent to the fracture for specimens strained parallel and perpendicular to ED are shown in Fig. \ref{fig4}(e) and (f), respectively. Although both specimens exhibit heavily deformed and fragmented microstructures near failure, clear differences emerge in the active deformation mechanisms and associated \textcolor{black}{evolution of the texture}.
For loading parallel to ED, the EBSD maps reveal extensive intragranular orientation gradients and a high density of deformation twins embedded within a slip-dominated matrix, indicating cooperative slip-twin interaction during the late stages of deformation. Correspondingly, the pole figures show only moderate texture modification, consistent with progressive lattice rotation dominated by dislocation slip. In contrast, the specimen strained perpendicularly to ED exhibits more pronounced lattice reorientation, reflected by a strong redistribution of basal poles in the pole figures. This evolution is characteristic of twinning-dominated deformation, where thick $\{10\bar{1}2\}\langle10\bar{1}1\rangle$ tensile twins (misorientation $\approx\qty{86}{\degree}$) significantly reorient large portions, and in some cases entire grains, as reported previously \cite{brown2005,el2015}. In both loading orientations, fine compression twins $\{10\bar{1}1\}\langle10\bar{1}2\rangle$ are observed near fracture, likely formed under the complex stress state in the necking region, and locally promoting secondary tensile twinning and double-twin $\{10\bar{1}1\}$–$\{10\bar{1}2\}$ formation \cite{cizek2008,lentz2016}.
\textcolor{black}{The} microstructural observations are consistent with the \textcolor{black}{evolution of the stored energy} shown in Fig. \ref{fig3}. Slip-dominated deformation produces pronounced intragranular misorientation and grain fragmentation, indicative of progressive defect accumulation during sustained plastic flow and associated gradual energy dissipation. In contrast, twinning-dominated deformation is characterized by thick twin lamellae and significant lattice reorientation with limited intragranular refinement prior to failure. These structural differences support the measured partitioning behavior, where slip promotes distributed defect storage and dissipation, whereas twinning initially accommodates plastic work through discrete twin formation and enhanced energy storage.

Plastic deformation of extruded AZ31B magnesium alloy exhibits fundamentally different mechanical responses depending on the dominant deformation mechanism. Slip-dominated deformation results in stable plastic flow, with approximately half of the plastic work dissipated as heat. \textcolor{black}{In contrast,}  twinning-dominated deformation initially stores a larger fraction of \textcolor{black}{the} plastic work, leading to rapid strain hardening, early strain localization, and premature fracture. Microstructural observations confirm distinct defect evolution pathways associated with slip and twinning.
These results indicate that in HCP magnesium alloys, the Taylor-Quinney coefficient must be interpreted in the context of the active deformation mechanism \textcolor{black}{rather than treated as a material constant}. The competition between slip and twinning governs not only anisotropic hardening but also the balance between energy storage and dissipation, thereby controlling mechanical stability.

\section*{Acknowledgements}
This work was partially supported by the Polish National Science Center under Grant No. 2021/41/B/ST8/03345.\\The authors gratefully acknowledge Dr. Tomasz Płoci{\'n}ski for preparing the specimens for EBSD analysis and Prof. Maciej Szczerba for valuable discussions that contributed to the development of this paper.

\section*{Data Availability}
The data supporting the findings of this study are
openly available in Zenodo at https://doi.org/10.5281/zenodo.18220980.

\section*{Competing Interests}
The authors declare that they have no known competing financial interests or personal relationships that could have appeared to influence the work reported in
this paper.

\bibliography{bib_az31b_slip_twin}
\bibliographystyle{plain}

\end{document}